\documentstyle[aps]{revtex}
\begin{document}
\title{Temperature dependences of resistivity and magnetoresistivity for
half-metallic ferromagnets}
\author{V. Yu. Irkhin$^{*1}$ and M. I. Katsnelson$^{1,2}$}
\address{$^{1}$Institute of Metal Physics, 620219, Ekaterinburg, Russia \\
$^2$Department of Physics, Uppsala University, Box 530, SE-751 21 Uppsala,\\
Sweden}
\maketitle

\begin{abstract}
Peculiarities of transport properties of three- and two-dimensional
half-metallic ferromagnets are investigated, which are connected with the
absence of spin-flip scattering processes. The temperature and magnetic
field dependences of resistivity in various regimes are calculated. The
resistivity is proportional to $T^{9/2}$ for $T<T^{*}$ and to $T^{7/2}$ for $%
T>T^{*},$ $T^{*}$ being the crossover temperature for longitudinal
scattering processes. The latter scale plays also an important role in
magnetoresistance$.$ The contribution of non-quasiparticle (incoherent)
states to the transport properties is discussed. It is shown that they can
dominate in the temperature dependence of the impurity-induced resistivity
and in the tunnel junction conductivity.
\end{abstract}


\section{Introduction}

Half-metallic ferromagnets (HMF) \cite{degroot,IK,pickett} attract now a
growing attention of researchers, first of all, because of their importance
for ``spintronics'', or spin-dependent electronics \cite{prinz}. HMF have
metallic electronic structure for one spin projection (majority- or
minority-spin states), but for the opposite spin direction the Fermi level
lies in the energy gap \cite{degroot}. Therefore the corresponding
contributions to electronic transport properties have different orders of
magnitude, which can result in a huge magnetoresistance for heterostructures
containing HMF \cite{IK}. A discovery of the half-metallic ferromagnetism in
colossal magnetoresistance (CMR) materials like La$_{1-x}$Sr$_x$MnO$_3$ \cite
{park} has increased considerably the interest in this topic. Transport
properties of HMF are the subject of numerous experimental investigations
(see, e.g., recent works for CrO$_2$ \cite{CrO2} and NiMnSb \cite{NiMnSb},
and the reviews \cite{IK,ziese,nagaev}). At the same time, the theoretical
interpretation of these results is still a problem.

As for electronic scattering mechanisms, the most important difference
between HMF and ``standard'' itinerant electron ferromagnets like iron or
nickel is the absence of one-magnon scattering processes in the former case 
\cite{IK}. Two-magnon scattering processes have been considered many years
ago for both the broad-band case (weak $s-d$ exchange interaction) \cite{Ros}
and narrow-band case (``double exchange model'') \cite{ohata}. The obtained
temperature dependence of the resistivity have the form $T^{7/2}$ and $%
T^{9/2}$, respectively. At low enough temperatures the first result fails
and should be replaced by $T^{9/2}$ as well \cite{lutovinov}; the reason is
the compensation of the transverse and longitudinal contributions in the
long-wavelength limit which is a consequence of the rotational symmetry of
the $s-d$ exchange Hamiltonian \cite{nagaev1,AKI}.

Up to now there are no results which describe in the whole temperature
region the resistivity of HFM and especially its magnetic-field dependence
which is most interesting from the experimental point of view. Such
expressions are obtained in Sect.2. Apart from three-dimensional case
studied before \cite{Ros,ohata,lutovinov} we consider also two-dimensional
HMF keeping in mind, e.g., layered CMR compounds like LaSr$_2$Mn$_2$O$_7$
(for a review see Ref.\cite{nagaev}) which are almost half-metallic
according to the recent band-structure calculations \cite{deboer}.

Owing to peculiar band structure of HFM, an important role belongs to
incoherent (non-quasiparticle) states which occur near the Fermi level
because of correlation effects \cite{IK}. In Sect.3 we treat the
corresponding contributions to resistivity and discuss tunneling phenomena
in HMF.

\section{Resistivity and magnetoresistivity}

To investigate effects of interaction of current carriers with local moments
we use the Hamiltonian of the $s-d(f)$ exchange model 
\begin{equation}
{\cal H}=\sum_{{\bf k}\sigma }t_{{\bf k}}c_{{\bf k}\sigma }^{\dagger }c_{%
{\bf k}\sigma }-\sum_{{\bf qk}}I_{{\bf k,k+q}}\sum_{\alpha \beta }{\bf S_q}%
c_{{\bf k}\alpha }^{\dagger }\mbox {\boldmath $\sigma $}_{\alpha \beta }c_{%
{\bf k-q}\beta }-\sum_{{\bf q}}J_{{\bf q}}{\bf S}_{{\bf q}}{\bf S}_{-{\bf q}}
\label{H}
\end{equation}
where $c_{{\bf k}\sigma }^{\dagger }$, $c_{{\bf k}\sigma }$ and ${\bf S}_{%
{\bf q}}$ are operators for conduction electrons and localized spins in the
quasimomentum representation, the electron spectrum $t_{{\bf k}}$ is
referred to the Fermi level $E_F$, $I_{{\bf k,k+q}}$ is the $s-d(f)$
exchange parameter which will be put for simplicity ${\bf k}$-independent, $%
{\bf \sigma }$ are the Pauli matrices. In the spin-wave region we have 
\begin{equation}
{\cal H}={\cal H}_0-I(2S)^{1/2}\sum_{{\bf kq}}(c_{{\bf k}\uparrow }^{\dagger
}c_{{\bf k}+{\bf q}\downarrow }b_{{\bf q}}^{\dagger }+h.c.)+I\sum_{{\bf kqp}%
\sigma }\sigma c_{{\bf k}\sigma }^{\dagger }c_{{\bf k+q-p}\sigma }b_{{\bf q}%
}^{\dagger }b_{{\bf p}}^{}
\end{equation}
The zero-order Hamiltonian includes non-interacting electrons and magnons, 
\begin{eqnarray}
{\cal H}_0 &=&\sum_{{\bf k}\sigma }E_{{\bf k}\sigma }c_{{\bf k}\sigma
}^{\dagger }c_{{\bf k}\sigma }+\sum_{{\bf q}}\omega _{{\bf q}}b_{{\bf q}%
}^{\dagger }b_{{\bf q}}^{}, \\
E_{{\bf k}\sigma } &=&t_{{\bf k}}-\sigma \Delta /2,\omega _{{\bf q}%
}=2S(J_0-J_{{\bf q}}),
\end{eqnarray}
with $\Delta =2IS$ being the spin splitting which is included in ${\cal H}_0$%
, $b_{{\bf q}}^{\dagger },b_{{\bf q}}$ are the Holstein-Primakoff boson
operators. In the half-metallic case the spin-flip processes do not work in
the second order in $I$ since the states with one spin projection only are
present at the Fermi level. At the same time, we have to consider the
renormalization of the longitudinal processes in higher orders in $I$
(formally, we have to include the terms up to the second order in the
quasiclassical small parameter $1/S$). To this end we eliminate from the
Hamiltonian the terms which are linear in the magnon operators by using the
canonical transformation \cite{nagaev1}, $\widetilde{{\cal H}}=e^U{\cal H}%
e^{-U}$ with 
\begin{equation}
U=-I(2S)^{1/2}\sum_{{\bf kq}}\frac{c_{{\bf k}\uparrow }^{\dagger }c_{{\bf k}+%
{\bf q}\downarrow }b_{{\bf q}}^{\dagger }}{t_{{\bf k+q}}-t_{{\bf k}}+\Delta }%
-h.c.
\end{equation}
Then we obtain the effective Hamiltonian 
\begin{equation}
\widetilde{{\cal H}}={\cal H}_0+\frac 12\sum_{{\bf kqp}\sigma }({\cal A}_{%
{\bf kq}}^\sigma +{\cal A}_{{\bf k+q-p,q}}^\sigma )c_{{\bf k}\sigma
}^{\dagger }c_{{\bf k+q-p}\sigma }b_{{\bf q}}^{\dagger }b_{{\bf p}}^{}
\label{hef}
\end{equation}
Here

\begin{equation}
{\cal A}_{{\bf kq}}^\sigma =\sigma I\frac{t_{{\bf k+q}}-t_{{\bf k}}}{t_{{\bf %
k+q}}-t_{{\bf k}}+\sigma \Delta }  \label{amp}
\end{equation}
is the $s-d$ scattering amplitude which vanishes at $q\rightarrow 0$ and
thereby takes properly into account the rotational symmetry of
electron-magnon interaction. More general interpolation expression for the
effective amplitude which does not assume the smallness of $|I|$ or $1/S$
was obtained in Ref.\cite{AKI} by a variational approach; it does not differ
qualitatively from simple expression (\ref{amp}). In the case of a
considerably ${\bf k}$-dependent exchange parameter, which may be relevant
for real itinerant magnets including HFM, one has in (\ref{hef})

\begin{equation}
{\cal A}_{{\bf kq}}^\sigma \rightarrow {\cal A}_{{\bf kqp}}^\sigma =\sigma
I_{{\bf k,k+q-p}}-\frac{2I_{{\bf k,k+q}}^2S}{t_{{\bf k+q}}-t_{{\bf k}%
}+\sigma S(I_{{\bf k+q,k+q}}+I_{{\bf k,k}})}  \label{amp1}
\end{equation}

The most general and rigorous method for calculating the transport
relaxation time is the use of the Kubo formula for the conductivity $\sigma
_{xx}$ \cite{kubo}

\begin{equation}
\sigma _{xx}=\beta \int_0^\beta d\lambda \int_0^\infty dt\exp (-\varepsilon
t)\langle j_x(t+i\lambda )j_x\rangle  \label{kru}
\end{equation}
where $\beta =1/T,$ $\varepsilon \rightarrow 0,$

\[
{\bf j}=-e\sum_{{\bf k}\sigma }{\bf v}_{{\bf k}\sigma }c_{{\bf k}\sigma
}^{\dagger }c_{{\bf k}\sigma } 
\]
is the current operator, ${\bf v}_{{\bf k}\sigma }=\partial E_{{\bf k}\sigma
}/\partial {\bf k\ }$ is the electron velocity. Representing the total
Hamiltonian in the form ${\cal H}={\cal H}_0+{\cal H}^{\prime }$, the
correlator in (\ref{kru}) may be expanded in the perturbation ${\cal H}%
^{\prime }$ \cite{Nak}. In the second order we obtain for the electrical
resistivity 
\begin{equation}
\rho _{xx}=\sigma _{xx}^{-1}=\frac T{\langle j_x^2\rangle ^2}\int_0^\infty
dt\langle [j_x,{\cal H}^{^{\prime }}(t)][{\cal H}^{^{\prime }},j_x]\rangle
\end{equation}
where ${\cal H}^{\prime }(t)$ is calculated with the Hamiltonian ${\cal H}_0$%
. Provided that the perturbation Hamiltonian has the form

\begin{equation}
{\cal H}^{\prime }=\sum_{{\bf kk}^{\prime }\sigma \sigma ^{\prime }}\widehat{%
W}_{{\bf kk}^{\prime }}^{\sigma \sigma ^{\prime }}c_{{\bf k}\sigma
}^{\dagger }c_{{\bf k}^{\prime }\sigma ^{\prime }}
\end{equation}
we obtain

\begin{equation}
\rho _{xx}=\frac T{2\langle j_x^2\rangle ^2}e^2\sum_{{\bf kk}^{\prime
}\sigma \sigma ^{\prime }}(v_{{\bf k}\sigma }^x-v_{{\bf k}^{\prime }\sigma
^{\prime }}^x)^2\int_{-\infty }^\infty dt\langle \widehat{W}_{{\bf kk}%
^{\prime }}^{\sigma \sigma ^{\prime }}(t)\widehat{W}_{{\bf k}^{\prime }{\bf k%
}}^{\sigma ^{\prime }\sigma }\rangle \exp [i(E_{{\bf k}\sigma }-E_{{\bf k}%
^{\prime }\sigma ^{\prime }})t]  \label{rho}
\end{equation}
with

\[
\langle j_x^2\rangle =e^2\sum_{{\bf k}\sigma }(v_{{\bf k}}^x)^2n_{{\bf k}%
\sigma }(1-n_{{\bf k}\sigma }) 
\]
This approach is equivalent to the solution of the Boltzmann transport
equation by the variational method \cite{ziman}.

In the HFM situation the band states with one spin projection only, $\sigma =%
{\rm sign}I,$ are present at the Fermi level \cite{IK}. Below we consider
the case $I>0,$ $\sigma =+$ and omit the spin indices in the electron
spectrum. We find from Eq.(\ref{rho}) the following expression for the
transport relaxation time $\tau $ defined by $\sigma _{xx}=e^2\langle
(v^x)^2\rangle \tau $%
\begin{eqnarray}
\frac 1\tau &=&\frac \pi {4T}\sum_{{\bf kk}^{\prime }{\bf q}}(v_{{\bf k}%
}^x-v_{{\bf k}^{\prime }}^x)^2({\cal A}_{{\bf kq}}^{\uparrow }+{\cal A}_{%
{\bf k}^{\prime }{\bf ,q-k}^{\prime }{\bf +k}}^{\uparrow })^2N_{{\bf q}%
}(1+N_{{\bf q-k}^{\prime }+{\bf k}})  \label{tau} \\
&&\ \ \ \ \ \ \ \ \times n_{{\bf k}}(1-n_{{\bf k}^{\prime }})\delta (t_{{\bf %
k}^{\prime }}-t_{{\bf k}}-\omega _{{\bf q}}+\omega _{{\bf q-k}^{\prime }+%
{\bf k}})\left/ \sum_{{\bf k}}(v_{{\bf k}}^x)^2\delta (t_{{\bf k}})\right. 
\nonumber
\end{eqnarray}
where $N_{{\bf q}}$ and $n_{{\bf k}}$ are the Bose and Fermi functions. A
similar expression has been derived first in Ref.\cite{Ros}, but with the
replacement of the effective amplitude just by $I$. After some
transformations we obtain 
\begin{equation}
\frac 1\tau =\pi I^2\sum_{{\bf kpq}}(v_{{\bf k}}^x{\bf -}v_{{\bf k+q-p}%
}^x)^2\delta (t_{{\bf k}})\delta (t_{{\bf k+q-p}})(1+N_{{\bf q}})(1+N_{{\bf p%
}})\left( \frac{t_{{\bf k+q}}}{t_{{\bf k+q}}+\Delta }\right) ^2\frac{\beta
(\omega _{{\bf p}}-\omega _{{\bf q}})}{\exp \beta \omega _{{\bf p}}-\exp
\beta \omega _{{\bf q}}}\left/ \sum_{{\bf k}}(v_{{\bf k}}^x)^2\delta (t_{%
{\bf k}})\right.
\end{equation}
Averaging over the angles of the vector ${\bf k}$ leads to the result $%
1/\tau \propto I^2\Lambda $ with

\begin{equation}
\Lambda =\sum_{{\bf pq}}f_{{\bf pq}}\frac{\beta (\omega _{{\bf p}}-\omega _{%
{\bf q}})|{\bf p-q}|}{\exp \beta \omega _{{\bf p}}-\exp \beta \omega _{{\bf q%
}}}(1+N_{{\bf q}})(1+N_{{\bf p}})  \label{lam}
\end{equation}
where $f_{{\bf pq}}=1$ for $p,q\gg q_0$ and

\begin{equation}
f_{{\bf pq}}=\frac{[{\bf p\times q]}^2}{({\bf p-q)}^2q_0^2}\,\,\,(p,q\ll
q_0).
\end{equation}
The wavevector $q_0$ determines the boundary of the region where ${\bf q}$%
-dependence of the amplitude become important, so that $t({\bf k+q})-t({\bf k%
})\simeq \Delta $ at $q\simeq q_0$. In the case $q<q_0$ the simple
perturbation theory fails and we have to take into account the spin
splitting by careful collecting the terms of higher orders in $I$. In the
simple one-band model of HMF where $E_F<\Delta $ one has $q_0\sim \sqrt{%
\Delta /W}$ ($W$ is the conduction bandwidth, lattice constant is put to
unity) \cite{nagaev1}. Generally speaking, $q_0$ may be sufficiently small
provided that the energy gap is much smaller than $W$, which is the case for
real HMF systems. A ``crossover'' wavector may exist in principle even for
the narrow-band case (where, instead of spin splitting, the spin subbands
have different widths) provided that the Fermi level is close to the gap
edge for the spin projection $-\sigma $.

The quantity $q_0$ determines a characteristic temperature and energy scale 
\begin{equation}
T^{*}=Dq_0^2\propto D({\Delta /}W)  \label{T*fm}
\end{equation}
where $D\propto T_C/S$ is the spin-wave stiffness defined by $\omega _{{\bf %
q\rightarrow }0}=Dq^2$, $T_C$ is the Curie temperature. Note that in the
case of an usual ferromagnetic metal the scale for existence of one-magnon
processes is smaller, $T_1^{*}\propto D({\Delta }/W)^2$.

When estimating temperature dependences of resistivity one has to bear in
mind that each power of $p$ or $q$ yields $T^{1/2}.$ At very low
temperatures $T<T^{*}$ small quasimomenta $p,q<q_0$ yield main contribution
to the integrals. Averaging the quantity (\ref{lam}) over the angle between
the vectors ${\bf p}$ and ${\bf q}$ we derive 
\begin{equation}
\Lambda =\frac 8{15q_0^2}\sum_{{\bf pq}}(5p_{+}^2-p_{-}^2)\frac{p_{-}^2}{%
p_{+}}\frac{\beta (\omega _{{\bf p}}-\omega _{{\bf q}})}{\exp \beta \omega _{%
{\bf p}}-\exp \beta \omega _{{\bf q}}}(1+N_{{\bf q}})(1+N_{{\bf p}})
\end{equation}
with $p_{+}=\max (p,q),p_{-}=\min (p,q)$. Then we obtain for the resistivity 
\begin{equation}
\rho (T)\propto (T/T_C)^{9/2}
\end{equation}
Such a dependence was obtained in the narrow-band case (double-exchange
model with large $|I|$), where the scale $T^{*}$ is absent \cite{ohata}, and
by the diagram approach in the broad-band case \cite{lutovinov}. At the same
time, at $T>T^{*}$ the function $f_{{\bf pq}}$ in Eq. (\ref{lam}) can be
replaced by unity to obtain 
\begin{equation}
\rho (T)\propto (T/T_C)^{7/2}
\end{equation}
This result is in agreement with the old works \cite{Ros}.

Now we treat the two-dimensional ($2D$) situation which may be appropriate
for layered manganites like La$_{2-x}$Ca$_{1+x}$Mn$_2$O$_7$ \cite
{nagaev,deboer}. At low temperatures we obtain 
\begin{equation}
\rho (T<T^{*})\propto (T/T_C)^{7/2}
\end{equation}
At the same time, for $T>T^{*}$ we obtain after replacing the scattering
amplitude by unity a logarithmically divergent integral which should be cut
at $T^{*}$. Thus we get 
\begin{equation}
\rho (T>T^{*})\propto (T/T_C)^{5/2}\ln (T/T^{*})
\end{equation}

To calculate the magnetoresistivity we introduce the gap in the magnon
spectrum, $\omega _{{\bf q\rightarrow }0}=Dq^2+\omega _0.$ Provided that the
external magnetic field $H$ is large in comparison with the anisotropy gap, $%
\omega _0$ is proportional to $H$ . In the $3D$ case the resistivity at $%
T<T^{*}$ is linear in magnetic field, 
\begin{equation}
\rho (T,H)-\rho (T,0)\propto -\omega _0T^{7/2}/T_C^{9/2}
\end{equation}
The situation at $T>T^{*}$ is more interesting since the quantity 
\[
\frac{\partial \Lambda }{\partial \omega _0}\propto \sum_{{\bf q}}q\omega _{%
{\bf q}}N_{{\bf q}}(1+N_{{\bf q}})\sum_{{\bf p}}\frac 1{\omega _{{\bf p}}^2}%
\propto \left( \frac T{T_C}\right) ^3\sum_{{\bf p}}\frac 1{\omega _{{\bf p}%
}^2} 
\]
contains a divergence which is cut at $\omega _0$ or $T^{*}$. We have at $%
T>\omega _0,T^{*}$

\begin{equation}
\delta \rho (T,H)\propto -\frac{T^3\omega _0}{[\max (\omega _0,T^{*})]^{1/2}}
\end{equation}
(of course, at $T<\omega _0$ the resistivity is exponentially small). A
negative $H$-linear magnetoresistance was observed recently in CrO$_2$ \cite
{CrO2}.

In the $2D$ case we obtain

\begin{equation}
\frac{\partial \Lambda }{\partial \omega _0}\propto T^{5/2}\sum_{{\bf p}}%
\frac{\phi ({\bf p})}{\omega _{{\bf p}}^2}
\end{equation}
where $\phi (p\ll q_0)=p^2/q_0^2,\phi (p\gg q_0)=1.$ This integral diverges
logarithmically at $\omega _0\ll T^{*}$ and as $\omega _0^{-1}$ at $\omega
_0\gg T^{*}.$ Taking into account the lower limit cutoff we derive

\begin{eqnarray}
\delta \rho (T,\omega _0 &\ll &T^{*})\propto -\left( \frac T{T_C}\right)
^{5/2}\frac{\omega _0}{T^{*}}\ln \frac{T^{*}}{\omega _0}, \\
\delta \rho (T,\omega _0 &\gg &T^{*})\propto -\left( \frac T{T_C}\right)
^{5/2}\ln \frac{\omega _0}{\max (T,T^{*})}
\end{eqnarray}
We see that simple replacement of the electron-magnon scattering amplitude
by $I$ does not enable one to describe correctly magnetoresistance even at $%
H>T^{*}.$

\section{Non-quasiparticle contributions to transport properties}

Now we treat the impurity contributions to transport properties in the
presence of potential scattering (they were considered first in \cite{IKT},
see also \cite{IK}). To second order in the impurity potential $V$ we derive
for the electron Green's function 
\begin{equation}
G_{{\bf kk}^{\prime }\sigma }(E)=\delta _{{\bf kk}^{\prime }}G_{{\bf k}%
\sigma }^{(0)}(E)+VG_{{\bf k}\sigma }^{(0)}(E)G_{{\bf k}^{\prime }\sigma
}^{(0)}(E)[1+V\sum_{{\bf p}}G_{{\bf p}\sigma }^{(0)}(E)]
\end{equation}
where 
\begin{equation}
G_{{\bf k}\sigma }^{(0)}(E)=[E-E_{{\bf k}\sigma }-\Sigma _{{\bf k}\sigma
}(E)]^{-1}
\end{equation}
is the exact Green's function for the ideal crystal. In the second order in $%
I$ the electron self-energy has the form 
\begin{equation}
\Sigma _{{\bf k}\sigma }(E)=2I^2S\sum_{{\bf q}}\frac{f(\sigma E_{{\bf k+q,-}%
\sigma })+N_{{\bf q}}}{E-E_{{\bf k+q,-}\sigma }+\sigma \omega _{{\bf q}}}
\end{equation}
with $f(E)$ the Fermi function.

Neglecting vertex corrections and averaging over impurities we obtain for
the transport relaxation time 
\begin{equation}
\delta \tau _{imp}^{-1}(E)=-2V^2{\rm Im}\sum_{{\bf p}}G_{{\bf p}\sigma
}^{(0)}(E)
\end{equation}
Thus the contributions under consideration are determined by the energy
dependence of the density of states $N(E)$ for the interacting system near
the Fermi level. The most nontrivial dependence comes from the
non-quasiparticle (incoherent) states with the spin projection $-\sigma =-%
{\rm sign}I,$ which are present near $E_F$ (Fig.1). They originate from the
imaginary part of the electron self-energy \cite{edwards,AI,IK1,IK}. We
obtain at $T=0$ 
\begin{equation}
\delta N_{incoh}(E)=2I^2S\sum_{{\bf kq}}f(-\sigma E_{{\bf k+q,}\sigma
})\delta (E-E_{{\bf k+q,}\sigma }-\sigma \omega _{{\bf q}})/(E_{{\bf k+q,}%
\sigma }-E_{{\bf k,-}\sigma })^2  \label{inc}
\end{equation}
The contribution (\ref{inc}) is asymmetric and vanishes at $E_F$ (Figs.1,2).
Near the Fermi level it is determined by the magnon density of states $%
g(\omega )$ and follows a power law, 
\begin{equation}
\delta N_{incoh}(E)\propto \int_0^{\sigma E}d\omega g(\omega )\propto
|E|^\alpha \theta (\sigma E)\,\,(|E|\ll \overline{\omega }).
\end{equation}
Here $\overline{\omega }$ is the maximum magnon frequency, $\theta (x)$ is
the step function, $E$ is referred to $E_F$; we have $\alpha =3/2$ and $%
\alpha =1$ for $3D$ and $2D$ cases, respectively. The corresponding
correction to resistivity reads 
\begin{equation}
\frac{\delta \rho _{imp}(T)}{\rho ^2}=-\delta \sigma _{imp}(T)\propto
-V^2\int dE\left( -\frac{\partial f(E)}{\partial E}\right) \delta
N_{incoh}(E)\propto T^\alpha  \label{rimp}
\end{equation}
The contribution of the order of $T^\alpha $ with $\alpha \simeq 1.65$
(which is not too far from 3/2) has been observed recently in the
temperature dependence of the resistivity for NiMnSb \cite{NiMnSb}. The
incoherent contribution to magnetoresistivity is given by 
\begin{equation}
\delta \rho _{imp}(T,H)\propto \omega _0\partial \delta N_{incoh}(\sigma
T)/\partial T\propto \omega _0T^{\alpha -1},
\end{equation}
so that we obtain a temperature-independent term in the $2D$ case.

The non-quasiparticle states in HMF can be probed also by nuclear magnetic
resonance (NMR) since they lead to the unusual temperature dependence for
the longitudinal nuclear magnetic relaxation rate, $1/T_1\propto T^{5/2},$
instead of the $T$-linear Korringa contribution which is absent in HMF \cite
{IK1,NMR}. Another useful tool is provided by tunneling phenomena \cite{AI1}%
, especially by Andreev reflection spectroscopy for a HMF-superconductor
tunnel junction \cite{falko}. The most direct way is probably the
measurement of a tunnel current between two pieces of HMF with the opposite
magnetization directions. To this end we consider a standard tunneling
Hamiltonian (see, e.g., Ref. \cite{mahan})

\begin{equation}
{\cal H} ={\cal H}_L+{\cal H}_R+\sum_{{\bf kp}}(T_{{\bf kp}}c_{{\bf k}%
\uparrow }^{\dagger }c_{{\bf p}\downarrow }+h.c.)
\end{equation}
where ${\cal H}_{L,R}$ are the Hamiltonians of the left (right) half-spaces,
respectively, ${\bf k}$ and ${\bf p}$ are the corresponding quasimomenta,
and spin projections are defined with respect to the magnetization direction
of a given half-space (spin is supposed to be conserving in the ``global''
coordinate system). Carrying out standard calculations of the tunneling
current ${\cal I}$ in the second order in $T_{{\bf kp}}$ one has (cf. \cite
{mahan})

\begin{equation}
{\cal I}\propto \sum_{{\bf kqp}}|T_{{\bf kp}}|^2[1+N_{{\bf q}}-f(t_{{\bf p-q}%
})][f(t_{{\bf k}})-f(t_{{\bf k}}+eV)]\delta (eV+t_{{\bf k}}-t_{{\bf p-q}%
}+\omega _{{\bf q}})
\end{equation}
where $V$ is the bias voltage. For $T=0$ one obtains

\begin{equation}
d{\cal I}/dV\propto \delta N_{incoh}(eV).
\end{equation}

To conclude, we have considered peculiarities of transport properties of
half-metallic ferromagnets which are connected with the unusual electronic
structure of these systems. Further experimental investigations would be of
great importance, especially keeping in mind possible role of HMF for
applications \cite{IK,pickett,prinz}.

The research described was supported in part by Grant No.00-15-96544 from
the Russian Basic Research Foundation, by Russian Science Support
Foundation, and by the Netherlands Organization for Scientific Research
(grant NWO 047-008-16).

{\sc figure captions}

Fig.1. Density of states in a half-metallic ferromagnet with $I>0$.
Non-quasiparticle states with $\sigma =-$ are absent below the Fermi level

Fig.2. Density of states in a half-metallic ferromagnet with $I<0$.
Non-quasiparticle states with $\sigma =+$ occur below the Fermi level


\begin{references}
\bibitem[*]{D}  E-mail: Valentin.Irkhin@imp.uran.ru

\bibitem{degroot}  R. A. de Groot, F. M. Mueller, P. G. van Engen, and K. H.
J. Buschow, Phys. Rev. Lett. {\bf 50}, 2024 (1983).

\bibitem{IK}  V. Yu. Irkhin and M. I. Katsnelson, Uspekhi Fiz. Nauk {\bf 164}%
, 705 (1994) [Phys. Usp. {\bf 37}, 659 (1994)].

\bibitem{pickett}  W. E. Pickett and J. Moodera, Phys. Today {\bf 54}(5), 39
(2001).

\bibitem{prinz}  G. A. Prinz, Science {\bf 282}, 1660 (1998).

\bibitem{park}  J. H. Park, E. Vescovo, H. J. Kim, C. Kwon, R. Ramesh, and
T. Venkatesan, Nature {\bf 392}, 794 (1998).

\bibitem{CrO2}  M. Rabe, J. Pommer, K. Samm, B. Oezyilmaz, C. Koenig, M.
Fraune, U. Ruediger, G. Guentherodt, S. Senz, and D. Hesse, J. Phys.:
Condens. Matter {\bf 14}, 7 (2002).

\bibitem{NiMnSb}  C. N. Borca, T. Komesu, H.-K. Jeong, P. A. Dowben, D.
Ristoiu, Ch. Hordequin, J. P. Nozieres, J. Pierre, S. Stadler, and Y. U.
Idzerda, Phys. Rev. B {\bf 64}, 052409 (2001).

\bibitem{ziese}  M. Ziese, Rep. Prog. Phys. {\bf 65}, 143 (2002).

\bibitem{nagaev}  E. L. Nagaev, Phys. Rep. {\bf 346}, 388 (2001).

\bibitem{Ros}  M. Roesler, phys. stat. sol.{\bf 8}, K31 (1965); F.
Hartman-Boutron, Phys. Kond. Mat. {\bf 4}, 114 (1965).

\bibitem{ohata}  K. Kubo and N. Ohata, J. Phys. Soc. Japan {\bf 33}, 21
(1972).

\bibitem{lutovinov}  V. S. Lutovinov and M. Yu. Reizer, Zh. Eksp. Theor.
Fiz. {\bf 77}, 707 (1979).

\bibitem{nagaev1}  A. P. Grigin and E. L. Nagaev, phys. stat. sol. (b) {\bf %
61}, 65 (1974); E. L. Nagaev, {\it Physics of Magnetic Semiconductors} (Mir,
Moscow, 1983).

\bibitem{AKI}  M. I. Auslender, M. I. Katsnelson, and V. Yu. Irkhin, Physica
B {\bf 119}, 309 (1983).

\bibitem{deboer}  P. K. de Boer and R. A. de Groot, Phys. Rev. B {\bf 60},
10758 (1999).

\bibitem{kubo}  R. Kubo, J. Phys. Soc. Japan {\bf 12}, 570 (1957).

\bibitem{Nak}  H. Nakano, Prog. Theor. Phys. {\bf 17}, 145 (1957); H. Mori,
Prog. Theor. Phys.{\bf 34}, 399 (1965).

\bibitem{ziman}  J. M. Ziman, {\it Electrons and Phonons - The Theory of
Transport Phenomena in Solids} (Oxford Univ. Press, Oxford, 2001).

\bibitem{IKT}  V. Yu. Irkhin, M. I. Katsnelson, and A. V. Trefilov, Physica C%
{\bf 160}, 397 (1989); Zh. Eksp. Theor. Fiz. {\bf 105}, 1733 (1994) [Sov.
Phys. JETP {\bf 78}, 936 (1994)].

\bibitem{edwards}  D. M. Edwards and J. A. Hertz, J. Phys. F {\bf 3}, 2191
(1973).

\bibitem{AI}  M. I. Auslender and V. Yu. Irkhin, J. Phys. C {\bf 18}, 3533
(1985).

\bibitem{IK1}  V. Yu. Irkhin and M. I. Katsnelson, J. Phys. : Condens.
Matter {\bf 2}, 7151 (1990).

\bibitem{NMR}  V. Yu. Irkhin and M. I. Katsnelson, Eur. Phys. J. B {\bf 19},
401 (2001).

\bibitem{AI1}  M. I. Auslender and V. Yu. Irkhin, Sol. State Commun. {\bf 56}%
, 703 (1985).

\bibitem{falko}  G. Tkachov, E. McCann, and V. I. Fal'ko, Phys. Rev. B {\bf %
65}, 024519 (2001).

\bibitem{mahan}  G. D. Mahan, {\it Many-Particle Physics} (Plenum Press, New
York, 1990), Sect. 9.3.
\end{references}
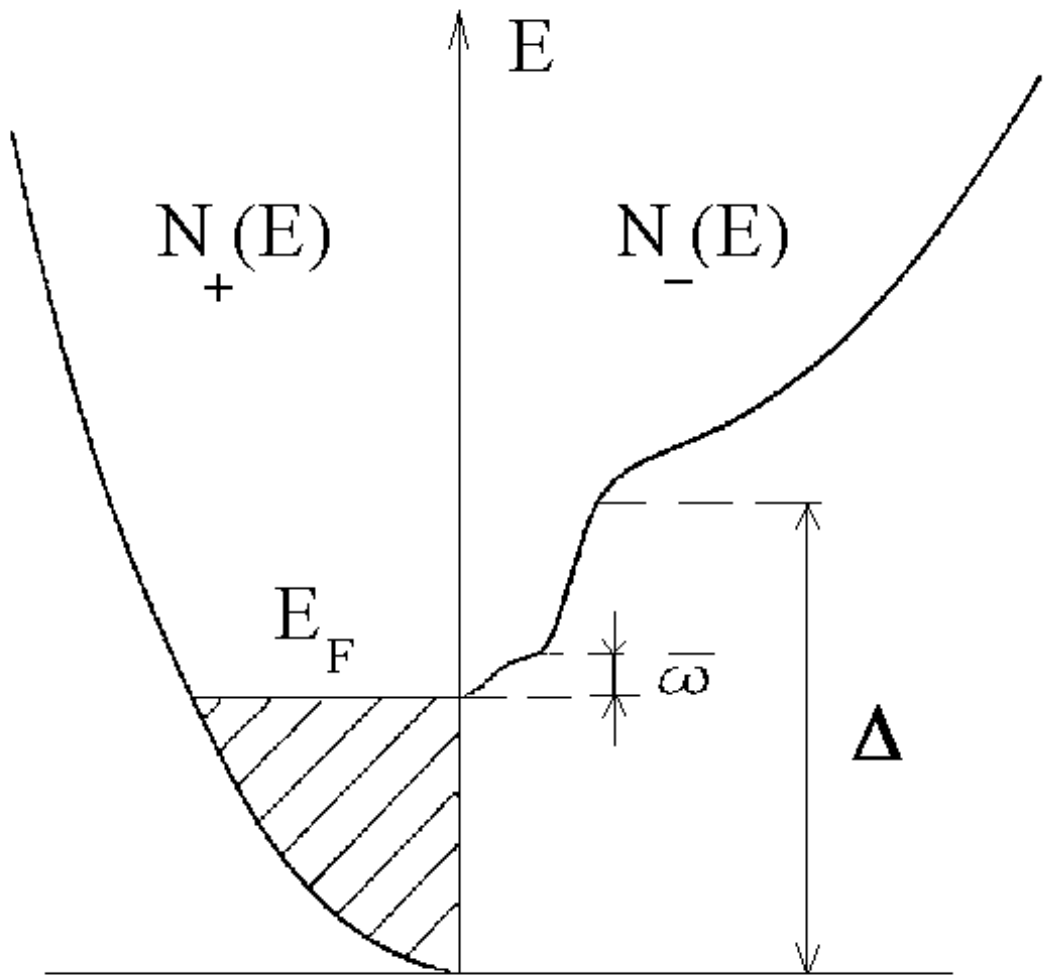
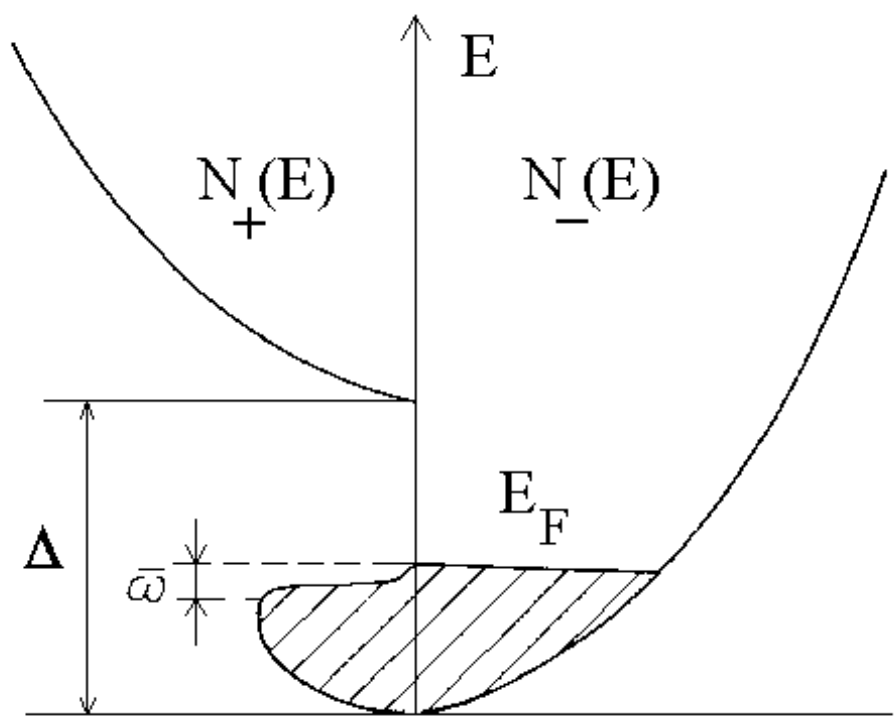
\end{document}